\documentclass[a4paper,conference]{IEEEtran}  

\IEEEoverridecommandlockouts
\usepackage[left=1.57cm,right=1.57cm,top=1.7cm,bottom=4.5cm]{geometry}
\def\BibTeX{{\rm B\kern-.05em{\sc i\kern-.025em b}\kern-.08em
    T\kern-.1667em\lower.7ex\hbox{E}\kern-.125emX}}
\usepackage[utf8]{inputenc}
\usepackage[backend=biber, style=ieee]{biblatex}
\usepackage{amsmath,amsfonts, amsthm}
\usepackage{algorithmic}
\usepackage{array}
\usepackage[caption=true,font=footnotesize,labelfont=sf,textfont=sf]{subfig}
\usepackage{tcolorbox}
\usepackage{color}
\usepackage{enumitem}
\usepackage{url}
\usepackage{graphicx}
\usepackage[hidelinks]{hyperref}
\usepackage{wrapfig}
\theoremstyle{definition}
\newtheorem{definition}{Definition}

\usepackage{orcidlink}
\usepackage[normalem]{ulem}
\usepackage{longtable, booktabs, etoolbox}
\usepackage{multirow}
\usepackage[noabbrev,nameinlink]{cleveref}
\bibliography{dt-smartfarming}  

\renewcommand{\itemautorefname}{\@gobble}
\usepackage{xspace}
\newcommand{\dt}{digital twin\xspace}
\newcommand{\dts}{digital twins\xspace}
\newcommand{\pt}{physical twin\xspace}

\newcommand{\ifz}{Industry 4.0\xspace}
\newcommand{\dtp}{Digital Twin Prototype\xspace}
\newcommand{\dtps}{Digital Twin Prototypes\xspace}
\newcommand{\gazebo}{GAZEBO\xspace}

\newcommand{\pubsub}{publish/subscribe\xspace}
\newcommand{\hil}{HIL\xspace}
\newcommand{\sil}{SIL\xspace}
\newcommand{\mil}{MIL\xspace}

\renewenvironment{quote}
  {\small\list{}{\leftmargin=0.45cm \rightmargin=0.25cm}%
   \item\relax}
  {\endlist}

\begin{document}

\title{Enabling Automated Integration Testing of Smart Farming Applications via Digital Twin Prototypes}

\author{
\IEEEauthorblockN{Alexander Barbie\IEEEauthorrefmark{1}, Wilhelm Hasselbring\IEEEauthorrefmark{1}, Malte Hansen\IEEEauthorrefmark{1}} \\ 
\IEEEauthorblockA{\IEEEauthorrefmark{1}Software Engineering Group, Christian-Albrechts-University, Kiel (Germany)}
}

\IEEEtitleabstractindextext{%
\begin{abstract}
\ifz represents a major technological shift that has the potential to transform the manufacturing industry, making it more efficient, productive, and sustainable. Smart farming is a concept that involves the use of advanced technologies to improve the efficiency and sustainability of agricultural practices. Industry 4.0 and smart farming are closely related, as many of the technologies used in smart farming are also used in \ifz.
Digital twins have the potential for cost-effective software development of such applications. 

With our \dtp approach, all sensor interfaces are integrated into the development process, and their inputs and outputs of the emulated hardware match those of the real hardware. The emulators respond to the same commands and return identically formatted data packages as their real counterparts, making the \dtp a valid source of a digital shadow, i.e. the \dtp is a prototype of the physical twin and can replace it for automated testing of the digital twin software. In this paper, we present a case study for employing our \dtp approach to automated testing of software for improving the making of silage with a smart farming application.

Besides automated testing with continuous integration, we also discuss continuous deployment of modular Docker containers in this context.
\end{abstract}

\begin{IEEEkeywords}
Smart Farming, Agricultural Machinery, Digital Twin Prototypes, Automated Testing, Continuous Integration
\end{IEEEkeywords}}

\maketitle

\IEEEdisplaynontitleabstractindextext

\section{Introduction}\label{sec:introduction}
Smart farming involves the use of sensors, IoT devices, and big data analytics.
Overall, the use of advanced technologies in both Industry 4.0 and smart farming has the potential to transform the way we produce food, making agriculture more efficient, sustainable, and resilient in the face of climate change and other challenges. For \ifz applications, the embedded software is becoming an increasingly crucial asset. With increasing requirements and, hence, increasing complexity, manufacturers require more engineers to manage the added complexity. The German \textcite{acatech} stated already in 2011 that ``there is a shortage of qualified engineers''. The demand for these engineers has dramatically increased globally at the beginning of this decade. While large software companies often develop software by distributed teams of engineers \cite{Jackson2022}, this is usually not the case for small and medium enterprises (SMEs) that develop embedded software systems. In  SMEs, embedded software is still often developed by the same engineers who also develop the electrical and/or mechanical parts. However, complex CPS require collaboration between mechanical, electrical, and software engineers \cite{acatech}.


With the increasing demand for context-aware, autonomous, and adaptive robotic systems \cite{DTroadmap}, more advanced software engineering methods need to be adopted by the embedded software community. As a result, the way we develop these systems must advance. In future development workflows, the embedded software systems will be the centerpiece of \ifz applications. To accomplish this, we need to shift from expert-centric tools \cite{DTroadmap} to modular systems that enable domain experts to contribute to different parts of the system.

Over the past decades, cost-saving techniques such as Model-in-the-Loop (\mil), Software-in-the-Loop (\sil), and Hardware-in-the-Loop (\hil) were introduced in the software development process of embedded software systems. They allow engineers to replace manual steps by utilizing software test beds for simulation and emulation. Although simulation and emulation help to reduce the costs for software development, \hil approaches are essential to validate and verify the embedded software system when sensors and actuators are involved~\cite{SILDemers2007}. To reduce the time embedded systems need to be validated manually by an engineer, we introduce the \dtp (DTP) approach \cite{MFI2020} that enables automated integration testing of embedded software system without the need of a physical connection to its hardware. 

This paper is structured as follows: In \Cref{sec:dtps}, we present the \dtp approach that allows the development and automated testing of \dts replacing the common \hil approach with \sil. \Cref{sec:relatedwork} discusses previous and related work with regards to \dts. SilageControl, the case under study, and the research design are described in \Cref{sec:silagecontrol} and \Cref{sec:researchdesign}. The results are presented and discussed in \Cref{sec:results}.  \Cref{sec:conclusion} draws our conlusions and takes a look at future work.

\section{Digital Twin Prototypes}\label{sec:dtps}
\textcite{dtdef-grieves} first used the term ``Digital Twin'' (DT) in a presentation for the establishment of a Product Lifecycle Management (PLM) center at the University of Michigan. Originating from the manufacturing industry in 2002, \dts became a buzzword in many marketing departments. In 2010, NASA introduced their vision of a \dt with a focus on modeling and simulation \cite{dtdef-nasa}. Meanwhile, digital twin research is a rapidly growing field, with applications in a wide range of industries. One of the central works around the understanding of \dts is by \textcite{dtdef-kritzinger}, who emphasized in their categorization of \dts that the physical object plays a crucial role in \dt development. The physical object provides the source for the digital shadow that synchronizes with the \dt. This is evident in \hil workflows, where engineers still need to test new code on the physical object, resulting in limited access and inconvenience for teams with multiple engineers. Asynchronous collaboration tools do not resolve the challenge of collaboration on embedded software systems, since engineers still need access to the hardware.

Today's modeling and simulation tools can immediately create a \dt of a single component or process, with \pubsub architectures allowing for all messages between processes to be captured and sent to a database or an IoT platform. However, complex \ifz applications require integrating multiple sensors and actuators into a larger system, posing a challenge with no simple solutions yet. The embedded community still uses various industrial interfaces and communication protocols such as ProfiBus, ProfiNet, ModBus, OpenCAN, OPC-UA, or MQTT, to name a few. Some of which are proprietary, making integration difficult.

Robust software testing for communication protocols is challenging due to the difficulty of emulating or simulating them. Software engineers often use mock-up functions in unit tests to avoid the exchange of data between processes, allowing them to obtain expected values. However, even robust unit testing with comprehensive edge case coverage is not enough. Therefore, some approaches use simulation tools that replace the communication protocols between hardware components with software interfaces. For \ifz applications, both approaches are inadequate, as insufficient testing can jeopardize the safety of human operators. Despite this, simulation tools are crucial for the development of \ifz applications as a source of data for sensors and actuators. We advocate an approach that involves emulating the hardware interface between a device driver and a sensor/actuator through virtualization and connecting the emulator to a simulation tool to obtain data from the simulated sensor/actuator. 
The device drivers can connect to either the real or emulated sensor/actuator without changing its configuration. Emulators can use either existing observed data or data from simulations, establishing a ground truth for the device driver without a physical connection to the real sensor/actuator. This virtualization also allows for the controls of the sensor/actuator to be used in automated tests or during the development process. We call this the \dtp approach. 
The \dtp takes the role of the \pt in the communication with the \dt, allowing to test the \dt software without access to a \pt. 

\begin{tcolorbox}
\begin{definition}[Digital Twin Prototype]
A Digital Twin Prototype (DTP) is the software prototype of a physical twin. The configurations are equal, yet the connected sensors/actuators are emulated. To simulate the behavior of the physical twin, the emulators use existing recordings of sensors and actuators. For continuous integration testing, the \dtp can be connected to its corresponding digital twin, without the availability of the physical twin.
\end{definition}
\end{tcolorbox}

Notice, that \textcite{dtdef-grieves} referred to a CAD model already as a \dtp. However, according to the categories by \textcite{dtdef-kritzinger}, this is only a digital model. 

The relationships between a \pt, \dt, and \dtp in our approach are depicted in Fig.~\ref{fig:dtpboxs}. The \dtp has the same software configuration as its corresponding \pt, but uses emulated sensors/actuators instead of real ones. The \dt can share parts of the software logic with the \pt, yet without connected sensors and actuators, and includes additional logic for controlling the \pt and \dtp. With this approach, all interfaces are integrated into the development process, and the inputs and outputs of the emulated hardware match those of the real hardware. The emulators respond to the same commands and return identically formatted data packages as their real counterparts, making the \dtp a valid source of a digital shadow, i.e. the \dtp is a prototype of the \pt and can replace it during development. 

\begin{figure}[b]
    \centering
    \includegraphics[width=.48\textwidth]{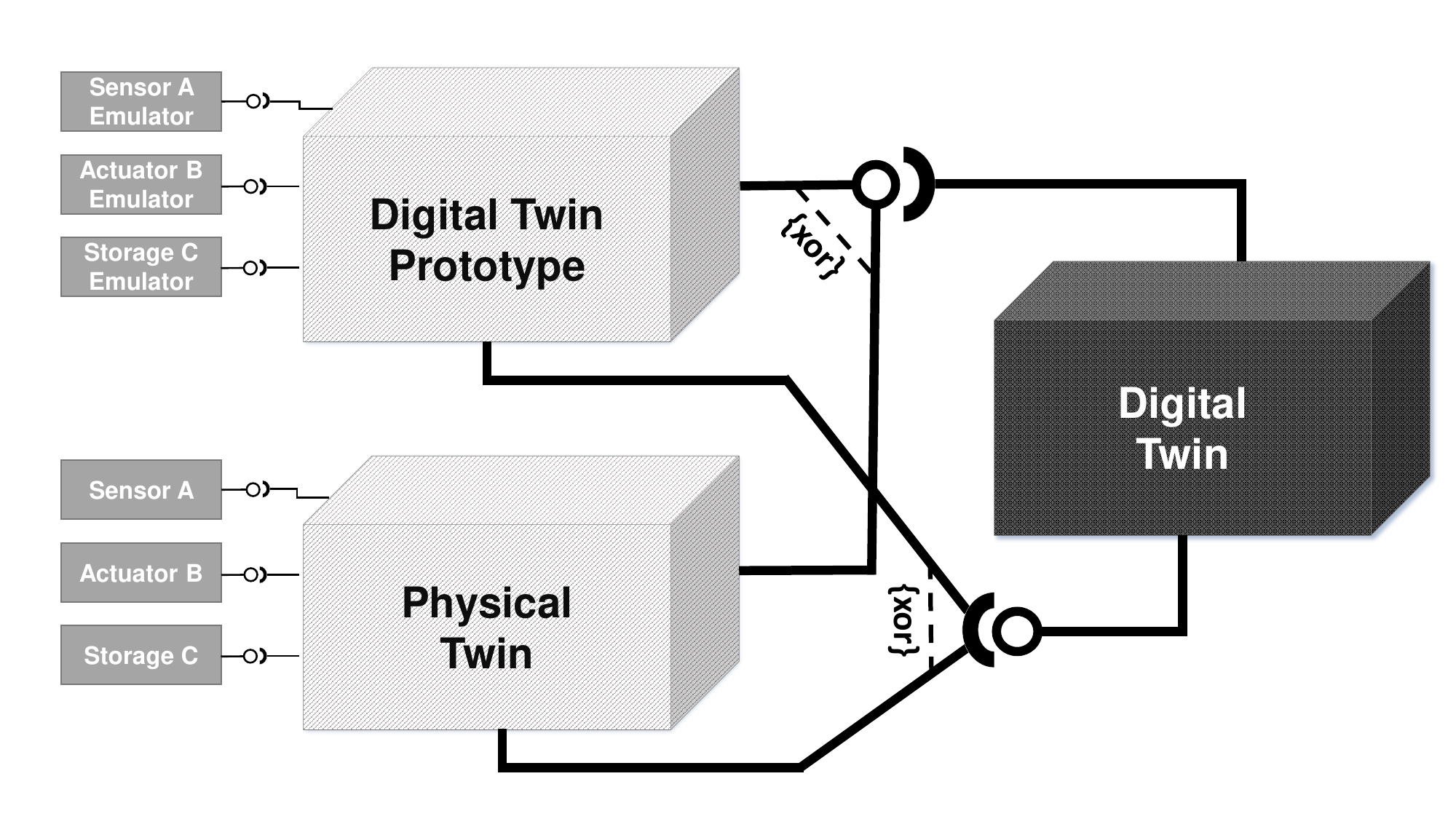}
    \caption{Relationships of Digital Twin Prototypes with physical twins and digital twins}
    \label{fig:dtpboxs}
\end{figure}

With \dtps, engineers can develop new software modules in their local IDE without a permanent connection to a physical test bed. However, before deployment to production, the entire system must still be tested on the \pt, as performance tests can only be conducted on hardware identical to that used in production. Nonetheless, \dtps allow for testing of the software logic independently of access to the hardware.

\section{Related and Previous Work}\label{sec:relatedwork}
Using robotic system for smart farming applications is not a new idea. \textcite{RaminShamshiri2018} give an overview of use cases, the state-of-the-art of agricultural robotics and the challenges faced. The overall challenges do not differ from that of other domains: digitalization, automation, and optimization. All robotic applications also face the general challenges inherited from the embedded software systems. 

The research of \dts in an agricultural context is still in the early stages. One reason could be, that the interpretation of \dts in the agricultural context may differ from the common interpretation and includes living things such as \dts of animals and crops~\cite{Pylianidis2021}, which is a completely different research field from robotics. Conversely, in our case study, the focus is solely on \dts of agricultural machines. \textcite{Pylianidis2021} conducted a study to investigate the added-value of \dts for agriculture. They found that agricultural \dts are still in a preliminary stage and are not designed thoroughly enough to compete with \dts in other disciplines. In their roadmap, they illustrate the evolution of agricultural \dts. In the first stage, a \dt includes monitoring, user interface, and analytic components. Next, actuators enrich the ability of \dts. In a third stage, simulations are included to support decision-making based on past and future predicted states of the \pt. This continuous including artificial intelligence and eventually, creating a \dt of the Earth.

In addition to the \dt definition by \textcite{dtdef-kritzinger}, we built on the \dt definition by \textcite{dtdef-saracco} and presented an overview of the approach in \textcite{MFI2020}.
The \dtp approach was developed and field tested in the project ARCHES (Autonomous Robotic Networks to Help Modern Societies). ARCHES was a Helmholtz Future Project with a consortium of partners from AWI (Alfred-Wegener-Institute Helmholtz Centre for Polar and Marine Research), DLR (German Aerospace Center), KIT (Karlsruhe Institute of Technology), and the GEOMAR (Helmholtz Centre for Ocean Research Kiel). Several \dtps for ocean observation systems were developed. The major aim of this project was to implement robotic sensor networks, which are able to autonomously respond to changes in the environment by adopting its measurement strategy, in both space and in the deep sea. 

This approach was evaluated during a research cruise, where a collaborative underwater network of ocean observation systems was established and deployed in the Baltic Sea. A field report on employing \dtps in this context is published by \textcite{demomission}. During that cruise, various scenarios were conducted to demonstrate the feasibility of digital twins for maritime environments. The realization of \dtps in ARCHES is illustrated in Fig.~\ref{fig:dtpreal}. With \dtps it was possible to develop and test scenarios before the mission took place. Automated testing was implemented through continuous integration / continuous delivery (CI/CD) in Gitlab. During the mission, all exchanged message on the \pt and \dt were recorded and can now be used to increase the quality of the CI/CD pipelines. 

\begin{figure}[hb]
    \centering
    \includegraphics[width=.48\textwidth]{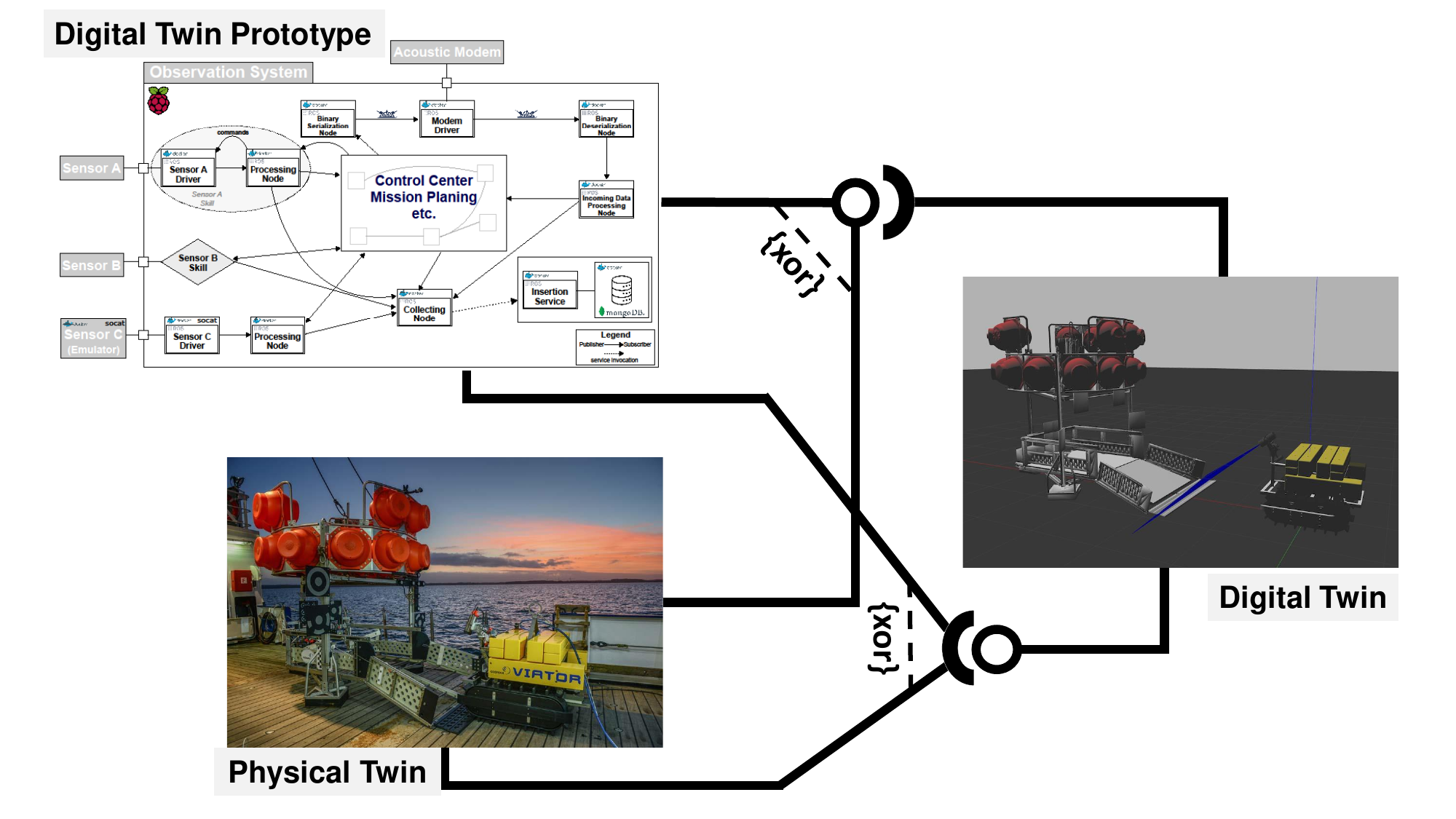}
    \caption{Realization and evaluation in ARCHES \cite{demomission}}
    \label{fig:dtpreal}
\end{figure}

The framework that was developed during the project ARCHES, called ARCHES Digital Twin Framework, is published open-source \cite{ADTF}. As the hardware of ocean observation systems is quite specific and not suitable for independent replication, we developed a \dtp of a PiCar-X by SunFounder \cite{sunfounder} and published this open-source \cite{abarbiegithub}, too. With the present paper, we extend this research by a case study in the smart farming domain.

\section{SilageControl}\label{sec:silagecontrol}
The smart farming project SilageControl with a consortium of the companies Silolytics GmbH, Blunk GmbH, and Kiel University aims to improve the process of silage making, i.e. the fermentation of grass or corn in silage heaps. To prevent the formation of mold, the harvested crop is compressed using heavyweight tractors.
As shown in Fig.~\ref{subfig:sctractor}, these tractors are equipped with a sensor bar (Fig.~\ref{subfig:scbar}) that includes GPS sensors, an inertial measurement unit (IMU), and a LiDAR. Together, these sensors provide continuous and accurate information about the tractor's position/orientation, and the shape and volume of the silage heap. Moreover, sensor data from modern harvesting machines may be added through telemetry to provide information about nutrient levels and provenance of each layer within the silage heap. The aggregated sensor data, external service data, and physics simulations are planned to be combined to compute the state of the silo. The first experiments were conducted using a Jetson Nano single-board computer by NVIDIA. Later, it was replaced by a Raspberry Pi.

\begin{figure}[t]
\centering
    \subfloat[Sensor bar in lab environment\label{subfig:scbar}]{%
      \includegraphics[width=0.45\textwidth]{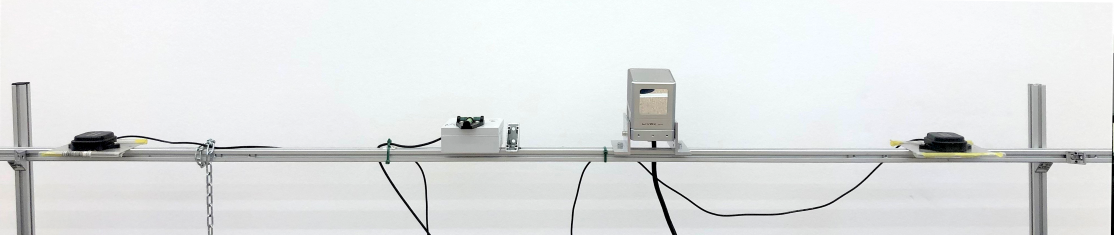} 
    }
    \\
    \subfloat[Sensor bar mounted on a tractor \label{subfig:sctractor}]{%
      \includegraphics[width=0.3\textwidth]{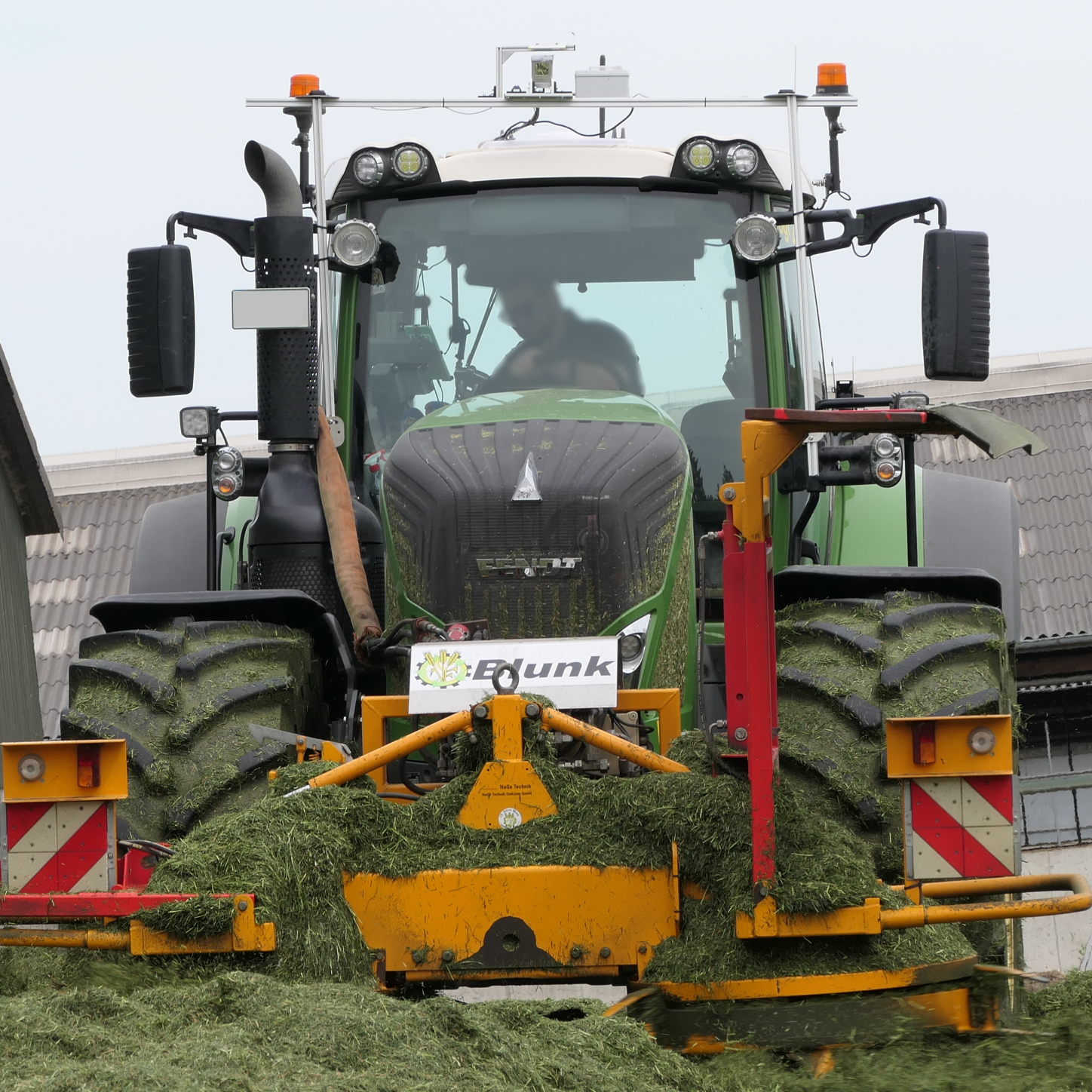} 
    }
    \caption{Sensor bar which monitors the process of silage making.}\label{fig:silagecontrolfg}
\end{figure}

The collected data provides near real-time insights into the volume of the silage heap, resulting in valuable feedback for the tractor driver to optimize the silo's compaction. However, since silage making is seasonal, the development and maintenance of the platform poses a significant challenge for SilageControl due to limited hardware availability during that time. Thus, the \dtp approach presents the prospect of improving the software development in this context.

\section{Research Design}\label{sec:researchdesign}
The software architecture of SilageControl is based on an event-driven microservice architecture with the Robot Operating System ROS~\cite{rospaper}, which enables us to conduct an exploratory case study that highlights the benefits of \dts in a real-world context. The goal of this study is as follows:

\begin{quote}
    \emph{\textbf{Goal:} Identify challenges faced with SilageControl that can be solved by the \dtp approach.}
\end{quote}

\noindent To identify the challenges faced by SilageControl, several meetings were held with Silolytics GmbH over the course of three years. For our case study, we report on the four most relevant meetings:
\begin{enumerate}
\item The first meeting took place in the summer of 2020. As the team was planning to develop the sensor system using ROS and had started with the idea at Kiel University, we invited them to the GEOMAR site and presented our \dtp approach, showing a live demonstration of ocean observation systems developed by us. There was no specific research objective for this meeting, as we only got to know each other and discussed the different roles, skills, and ideas in the team.

\item The second meeting was at the end of 2021 when we interviewed the lead engineer for an interview study. As nearly one year has passed, we were interested in their actual development process since the first meeting.
The questions we asked are presented in \Cref{table:questions}. When we elaborated these questions, we followed the guide by \textcite{carlaqualitativeresearch}. Note, that these questions are only the start questions. Since we conducted semi-structured interviews, they may vary from interview to interview depending on the conversation's flow. For example, we asked ``\emph{You mentioned you are developing XYZ, can you please describe how your development workflow looks like?}'' instead of ``\emph{Please describe a typical workflow, when you develop/adjust a feature?}''. Furthermore, depending the conversation's flow, follow-up questions may arise that are not listed here, yet are relevant to get to know their development processes.
\begin{table*}[ht]
\renewcommand{\arraystretch}{1.5}
\centering
\begin{tabular}{p{0.225\textwidth} | p{0.65\textwidth}}
\multicolumn{1}{c}{\textbf{Phase}} & \textbf{General Interview Questions}  \\ \toprule
\multicolumn{1}{c|}{Warm up}    & What is your companies' main product and which role do you have? \\ \midrule
\multicolumn{1}{c|}{\multirow{3}{*}{\shortstack{Preferred strategies and \\ observed obstacles}}}     & Please describe a typical workflow, when you develop/adjust a feature? \\ 
      & Do you remember a (complicated) problem you had to solve recently and how did you solve it?\\ 
          & What do you like about your approach and what do you dislike? \\ \midrule
\multicolumn{1}{c|}{\multirow{2}{*}{\shortstack{Quality assurance and \\ product customization}}} & How do you ensure that new features or bug fixes do not compromise your product? \\ 
          & How do you individualize the software of your product? \\
\bottomrule
\end{tabular}
\caption{The initial interview questions.}\label{table:questions}
\end{table*}

\item The third meeting took place in the spring of 2022, where we presented the ARCHES Digital Twin Framework and our \dtp using the PiCar-X example.
\item During the fourth meeting in January 2023, Silolytics presented their early adoption of \dtp.
\end{enumerate}

\section{Results and Discussion}\label{sec:results}
After the initial idea of SilageControl was formed, we invited the team to the GEOMAR Helmholtz Centre for Ocean Research in Kiel, where we demonstrated our research on ocean observation systems~\cite{demomission}. Since SilageControl is a modular sensor bar that is placed on top of a tractor and the team planned to utilize ROS for the software architecture, we expected to support the team by sharing our experience with embedded software development using ROS. During the meeting, we also invited a group of students whom we co-supervised in a master's project to test and demonstrate a web application on real ocean observation systems. The setup was already quite similar to the idea of SilageControl, where sensory data is visualized in a web application.

\subsection{The Challenges in the Development Workflow}
At the second meeting, we conducted an interview with the lead engineer to gain insight into the development workflow for integrating sensors into the SilageControl system. The lead engineer described the process as follows:

\begin{quote}
``\emph{When integrating a sensor, we always start by testing it on a PC. We identify which device drivers are required, install the drivers, and ensure they are compatible with the platform. We verify the sensor's outputs on the PC to confirm that they meet the expected results. Some software packages require specific coordinate systems, and testing [in the office] on the PC is the easiest way to determine compatibility. The next step is to install the sensor on the platform, using the hardware manual to guide integration and seeking input from the sales team if needed. We check the device drivers in the pipeline, building them in Docker for later deployment to the mobile platform, where we also use Docker for building.}''
\end{quote}

However, integrating sensors into the overall system can pose certain challenges. For example, integrating a GPS sensor into the office setting can be difficult due to:

\begin{quote}
    ``\emph{[...] One of our problems is that the GPS sensors need an unobstructed view of the sky, which complicates testing of the system. While indoor GPS systems are tested, we always face the challenge of setting up the sensors correctly when we have to take them outside, and it's difficult to maintain the required power supply. This issue is a bit of a sticking point, and we have yet to find a solution.}''
\end{quote}

As previously highlighted, embedded software development often faces the challenge of testing hardware in a suitable environment. Integrating a GPS sensor in an office setting is a typical example of this problem. The target system is too large for an office, and testing on a tractor in the field is expensive and time-consuming. To overcome this challenge, engineers often detach the hardware from the target system and place it in a laboratory environment. However, if a sensor only works as intended in the field, this problem becomes even more severe and can have a significant impact on the entire development process.

SilageControl conducted their first experiments in 2020 after visiting GEOMAR, and we also asked them about the lessons they learned from those experiments:

\begin{quote}
``\emph{We spent approximately 400 hours testing our system and collecting data from various use cases to cover as many scenarios as possible. Using this data, we developed our algorithms during the winter months, focusing primarily on perception, which involves recognizing the surrounding environment and identifying the building's structural features automatically. We then integrated the entire workflow into the app and tested it offline, which was challenging due to the lack of available data for testing. However, we recorded various use cases, which we can now use to test the system's functionality. We created launch files with [ROS] bag files to test our software and algorithms and evaluated them using different metrics. For instance, we assessed the importance of sensor alignment and the accuracy of LiDAR detection. We used this approach in our development workflow to ensure that our system meets the required standards.}''
\end{quote}

Acquiring realistic and reusable data from sensors and actuators is a challenging task for the development and testing of embedded software systems. To address this challenge, the team used their first trial period to gather as much data as possible from the sensors and actuators. They stored all the ROS messages sent through the system in ROS bag files, which can be replayed later. We further asked for positive and negative experience with their workflow:

\begin{quote}
``\emph{I really appreciate the automatic building of packages [with Docker]. It is quite straightforward to pull the packages onto the Jetson Nano later on. All you need to do is download the images. Additionally, GitHub has Docker registries where you can easily deploy your packages. I find it helpful that pushing to the latest branch results in the package being built automatically. If we were to do this manually, it could lead to problems, but with the base image, all dependencies are included. By shifting the problems, we only need to do it once in Docker, which is much more convenient than using SSH or connecting a monitor to pull everything onto the Jetson Nano. This is actually the main advantage, as you only have to do it once and not repeatedly.}''
\end{quote}

Regarding negative experiences, the engineer mentioned several challenges. First, some sensor drivers worked in older versions of ROS but not in the current LTS version. For instance, when the last LTS version (Noetic) for ROS, Version~1, was released in 2021, some of their drivers only supported the previous version (ROS Melodic). This required manual adjustment, which means that future updates might not be automatically loaded into the Docker container. Second, a LiDAR sensor malfunctioned after a few hours of use and costs several thousand Euros, which was significant in our context. Third, software documentation, especially for device drivers, is often unavailable or poorly maintained. The team faced the most difficulties with the LiDAR sensor, where they had to try several models before deciding on the right one.

Since the engineer mentioned Docker several times, we asked whether they had started using it before or after the meeting at GEOMAR and whether they used it for deployment/testing only or also for development. He said that they began using Docker for deployment and automated testing after meeting with us but had not used it for development beforehand.

\begin{quote}
``\emph{[...] When everything is in Docker containers, and you have to rebuild and change the images, you do not make as many changes anymore. It becomes more important that it works and is thoroughly tested. This also makes the entire deployment process more comprehensible.}''
\end{quote}

This reveals a misconception about the potential use of Docker in development. Docker actually allows the mounting of folders into a container, which further allows for easy integration of the ROS development environment into the container. This means that the ROS development environment can easily be mounted to the container and started from there, without having to rebuild the image and create a new container. We emphasized this point during the interview.

\subsection{The Challenges for Product Customization and Quality Assurance}
The third part of the interview was about the customization of SilageControl and the quality assurance workflow. As SilageControl is already built on a modular software architecture, changing sensors and their drivers is thought along the development of the system. Moreover, it is difficult to properly mount the sensor bar, since tractors and other argicultural machines are not standardized and different manufacturers exists. With regard to the customization of the sensor bar, SilageControl learned from their first experiments:

\begin{quote}
    ``\emph{[During our first trials], we had to [manually] position the system [(the single sensors)] precisely to the tractor during installation, which made it peculiar and inconvenient. [Since then], we improved the installation and introduced a rail containing the sensors that can now be mounted on the tractor in various ways. Additionally, we have standardized the rail's structure, which allows for calibration routines to run beforehand, and the sensors to calibrate themselves automatically. As a result, there is no need for pinpoint accuracy during installation. Before, it was nearly impossible to align everything perfectly. Now, our system is plug-and-play, making it easier to install. To ensure accuracy, we use CNC machines to manufacture the parts, including brackets and mounts. [...] Even a slight deviation of just one degree during installation can result in a significant error of 17$\,$cm over a meter, which is far from ideal [...]. [Thus] we are constantly working to improve the installation process and reduce the margin of error to achieve the highest level of precision [and accuracy] possible.}''
\end{quote}

Since the engineer already mentioned that the source code is managed on GitHub and the build process and unit tests are also executed in GitHub CI/CD pipelines, we did not ask whether there were automated unit tests, but rather whether there were automated integration tests that included the hardware in the CI/CD pipeline. This is particularly challenging for SMEs with limited funding for spare hardware. The engineers develop on the same hardware that is used in production, so it cannot be connected to CI/CD systems such as GitHub. The engineer confirmed this limitation.

\begin{quote}
``\emph{In my opinion, connecting all the sensors and having a setup to test all the basic functionalities using HIL with synthetic data would be beneficial. This would be relatively easy to implement, and for GPS, using synthetic data would be the simplest option. By doing so, we could test various forms of noise and other aspects. This approach has been well-researched for the sensor type, and it may help us resolve our problem. Although we have considered it, we have not found a solution yet, and we also need to consider the effort.}''
\end{quote}

During the silage season, the hardware is rarely available for the developers, which makes it difficult to connect the sensors to GitHub or perform automated integration tests. Since these sensors can be quite expensive, the team does not have any spare sensors to use during this time. Therefore, the team has already considered using simulations to simulate the tractor and sensor bar, and to obtain a virtual context. This idea was further explained when the engineer was asked about the quality assurance workflow for SilageControl:

\begin{quote}
``\emph{We intend to use simulations extensively in the future since we have other product development aspects that can only be developed in the simulation and would be too time-consuming otherwise. This is especially true now that we are collaborating with a service provider, and we intend to incorporate the \dt approach. This is also what we hope to accomplish here with Kiel University in the future.
We have already devised a plan for what we intend to change in the future, but we have never had the necessary environment to implement it. Therefore, this will be the primary work package in the near future.}''
\end{quote}

Thus, the Silolytics GmbH already identified possible approaches to solve their problems with costly and often unavailable hardware components.

\subsection{A \dtp for SilageControl}
\begin{figure}[b]
    \centering
    \includegraphics[width=.45\textwidth]{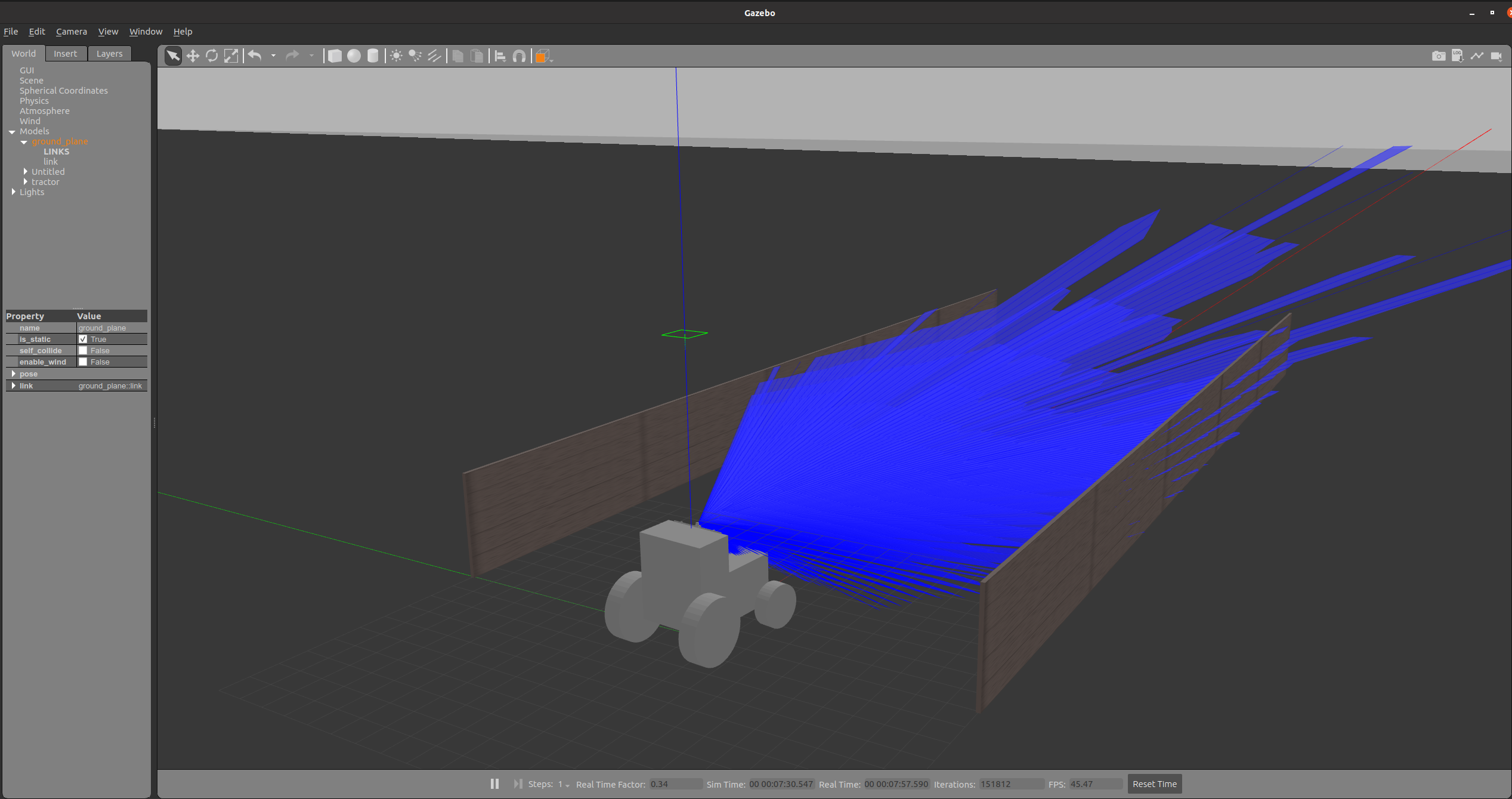}
    \caption{A simple digital model of a tractor mounted with the SilageControl sensor bar in a \gazebo simulation.}
    \label{fig:gazebosilolytics}
\end{figure}

\begin{figure*}[ht]
    \centering
    \includegraphics[width=.8\textwidth]{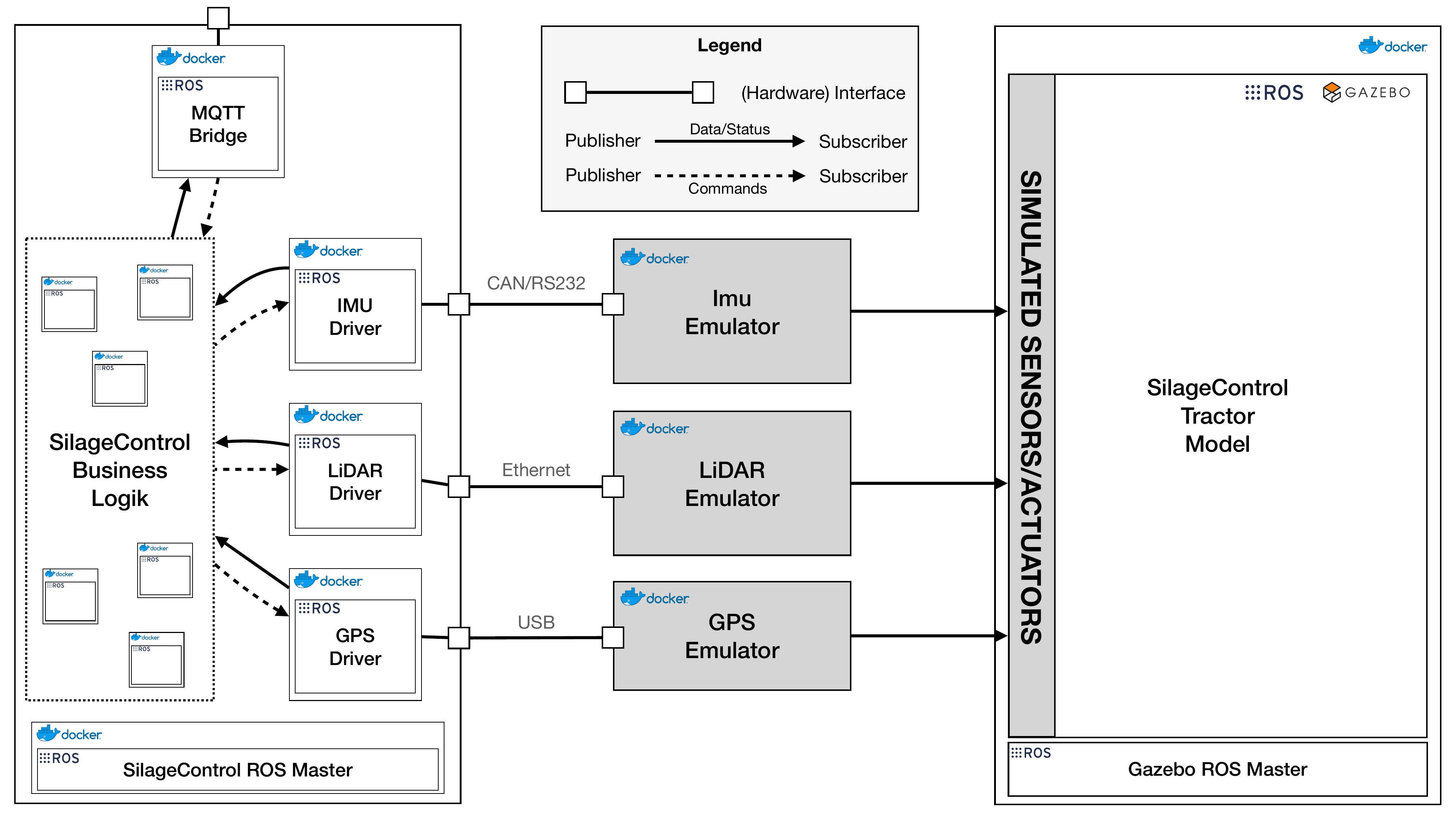}
    \caption{\dtp of a tractor mounted with the SilageControl sensor bar in a \gazebo simulation.}
    \label{fig:gazebosetupsilolytics}
\end{figure*}
During our third meeting, we presented the PiCar-X example and demonstrated how simulation can be used to provide virtual context to a \dtp. At the time of the meeting, the ARCHES Digital Twin Framework was not yet published and Silolytics already started developing a \dt using a combination of C++ and Python components. However, SilageControl started following the \dtp approach, and during our fourth meeting, they presented their first results for their \dt. The \gazebo simulation \cite{gazebopaper} was used to model a simple tractor including the sensors, as shown in \ref{fig:gazebosilolytics}. This model already contains the IMU, GPS and LiDAR sensors and the  business logic of SilageControl can be connected to the simulation. To replace the \pt with a \dtp for development and testing, the next step is to connect the device drivers with the emulators connected to the simulation as shown in Fig.~\ref{fig:gazebosetupsilolytics}. The emulators receive commands from the drivers via different interfaces (Ethernet, RS232, and USB) and send the commands to the GAZEBO simulation. The corresponding simulated sensors in \gazebo react to the sent command and return the simulated data to the emulator. The emulator forwards the data again to the corresponding device driver. 

The traditional approach would be to add the emulator logic as module to the simulation. However, this has a few drawbacks. Ethernet interfaces are easy to emulate in simulation tools. Other interfaces such as RS232 are not that easy to integrate to simulations and often need additional configuration. In addition, a tight coupling of an emulator to a specific simulation tool cannot be easily migrated to another simulation tool. Separating the emulator logic and only adding an interface to the simulation, allows an easier switch to other tools. Furthermore, SilageControl can decide to also use recorded data (e.g. from the ROS bag files) to replay the data to the emulator instead of using purely simulated data.

\subsection{Threads to Validity}\label{sec:threads}
We identified two threads to the validity of our case study. Firstly, the case study was not initially conducted with a strict research plan and the idea developed over time. Secondly, we had prior knowledge of the Silolytics GmbH and its team before conducting the interviews and already presented them our results from the ARCHES project. This might have influenced their development process and although the interviewed engineer did not know the questions beforehand, he may have guessed the intentions behind the different questions and answered them accordingly.

\section{Conclusion and Future Work}\label{sec:conclusion}
In summary, the challenges for the SilageControl project can be summarized as follows:
\begin{itemize}
    \item The hardware has to be developed on a test bed, but tested on a real target;
    \item The GPS sensors are difficult to test on the test bed;
    \item Data acquisition is only possible seasonal;
    \item Sensor prices are too high to buy spare hardware for development/testing;
    \item Available software modules are sometimes not well maintained;
\end{itemize}

As described in the introduction, these challenges are common and engineers face most of these challenges during the development of their systems. In the context of the project ARCHES, we described our \dtp approach that tackles these challenges~\cite{demomission}. For instance, virtualizing the hardware in the form of a \dtp and combining it with a simulation, as shown in our PiCar-X example, enables the usage of virtual context for development and testing. From the \dtp, the team also gets a \dt that can be utilized to monitor the \pt during operation and collect data without the need to physically connect to the sensor bar mounted on a tractor. This enables the development of embedded software systems without the need to physically connect to the hardware and hence reduces costs that may be needed for spare hardware otherwise. Without the need for hardware in the development loop, the seasonal data gathering missions become less of a problem.

Only the last point cannot be solved utilizing a \dt, as the software modules are external features. The problem at that point is the embedded software community that develops device drivers with tight coupling to the overall system, e.g. a specific middleware. However, as already discussed by \textcite{Kaupp2007} and by \textcite{rospaper}, device drivers should be independent of the middleware used in the development of the embedded system. Otherwise, when switching to a different middleware, the driver must be redeveloped, even if the logic of the hardware remains unchanged. Due to the vast amount of programming languages and middlewares, manufacturers of sensors and actuators cannot develop and maintain device drivers for all of them. Thus, it should be in the interest of manufacturers that their sensors are easy to access and independent of the middleware that is used. As the engineer explained, they tested several sensors, including the software, before deciding on one.

\section*{Acknowledgment}
This research is funded by the Federal Ministry of Food and Agriculture (BMEL, Germany) via the Federal Office for Agriculture and Food (BLE, Germany) in the SilageControl project 
(contract no. 281DT02B21).
 \printbibliography

\end{document}